\documentclass[useAMS,usenatbib]{mn2e}
\bibpunct{(}{)}{;}{a}{}{,}
\usepackage{graphicx}
\usepackage{latexsym}
\usepackage{multicol}

\title{Photoionising feedback in star cluster formation}

\author[J. E. Dale, I. A. Bonnell, C. J. Clarke, M. R. Bate]{J. E. Dale$^{1}$\thanks{E-mail: Jim.Dale@astro.le.ac.uk (JED)}, I. A. Bonnell$^{2}$, C. J. Clarke$^{3}$ and M. R. Bate$^{4}$\\
$^{1}$Department of Physics and Astronomy, University of Leicester, University Road, Leicester, LE1 7RH, UK\\
$^{2}$School of Physics and Astronomy, University of St Andrews, St Andrews, Fife, KY19 9AJ, UK\\
$^{3}$Institute of Astronomy, Madingley Road, Cambridge, CB3 0HA, UK\\
$^{4}$School of Physics, University of Exeter, Stocker Road, Exeter, EX4 4QL, UK}

\begin{document}

\pagerange{\pageref{firstpage}--\pageref{lastpage}} \pubyear{2002}

\maketitle

\label{firstpage}

\def\mnras{MNRAS}
\def\apj{ApJ}
\def\aap{A\&A}
\def\apjl{ApJL}
\def\apjs{ApJS}
\def\apss{ApSS}
\def\aj{AJ}

\begin{abstract}
We present the first ever hydrodynamic calculations of star cluster formation that
incorporate the effect of feedback from ionising radiation. In our simulations, the
ionising source forms in the cluster core at the intersection of several dense
filaments of inflowing gas. We show that these filaments collimate ionised outflows
and suggest such an environmental origin for at least some observed outflows
in regions of massive star formation. Our simulations show both positive
feedback (i.e. promotion of star formation in neutral gas compressed by expanding
HII regions) and negative feedback (i.e. suppression of the accretion
flow in to the central regions). We show that the volume filling factor of ionised
gas is very different in our simulations than would result from the case
where the central source interacted with an azimuthally smoothed gas density
distribution. As expected, gas density is the key parameter in determining
whether clusters are unbound by photoionising radiation. Nevertheless, we find
- on account of the acceleration of a small fraction of the gas to high
velocities in the outflows - that the deposition in the gas of an energy that
exceeds the binding energy of the cluster is {\it not} a sufficient criterion
for unbinding the bulk of the cluster mass.

\end{abstract}


\section{Introduction}

It has long been recognised that massive stars deposit considerable
mechanical and thermal energy into the ISM by a variety of processes.
On galactic scales, the main such feedback agent is supernova explosions
\citep[e.g.][]{1977ApJ...218..148M,1986ApJ...303...39D};
however the relatively long time lapse before supernovae explode - compared
with the dynamical timescale in star forming regions - ensures that
supernovae are ineffective in regulating star formation in the
close vicinity of the progenitor. At the opposite extreme of length
scales, radiation pressure on dust operates in the region ( a few hundred
A.U. from a typical OB star) where dust is close to its sublimation
temperature. This form of feedback is thus mainly instrumental
in controlling the formation of the OB star itself \citep{1986ApJ...310..207W,1987ApJ...319..850W,2002ApJ...569..846Y,2003MNRAS.338..962E, edgar04}.\\
\indent On intermediate scales ($\sim$ parsecs), stellar winds and photoionisation
are likely to play the primary role in regulating the formation of
star {\it clusters} \citep[e.g.][]{1986MNRAS.221..635T,1999MNRAS.309..332T,2003A&A...411..397T}. On the one
hand, the star formation rate may be enhanced by the fragmentation of
shells of dense gas swept up by winds/expanding HII regions
\citep[e.g.][]{1998orig.conf..150E,2002MNRAS.334..693E,2002MNRAS.329..641W}
but the acceleration and expulsion of gas from the neighbourhood of the
OB star also ultimately limits the fraction of a molecular cloud
core that can be turned into stars. This latter effect has important
implications for the formation of of bound star clusters, since -
by simple energetic arguments - a cluster cannot remain bound if it
rapidly loses more than $50 \%$ of its original mass \citep{1980ApJ...235..986H,1984ApJ...285..141L}. \citep[See also][for more sophisticated analyses of this problem]{1997MNRAS.284..785G,2000ApJ...542..964A,2001MNRAS.323..988G,2003MNRAS.338..673B}. Although the majority of stars form in 
clusters \citep{2000prpl.conf..151C}, most of these clusters are already
unbound at an age of $\sim 10^7$ years \citep{1991fesc.book..139B}.
It is likely, at least in clusters that contain more than a few
hundred members - and are thus populous enough to
contain an OB star - that such feedback effects are
decisive in unbinding nascent clusters.\\
\indent Feedback is also likely to be the main factor that regulates the
Galactic star formation rate (SFR). It has long been recognised that
the observed SFR in the Milky Way of a few M$_\odot$ yr$^{-1}$
is only  a few per cent  of the value that would result if all the Giant
Molecular Clouds (GMCs)  in the Galaxy were turning into stars on their
internal dynamical timescales \citep{1974ApJ...192L.149Z}. Although one solution is that star
formation in GMCs is quasistatic - i.e. is regulated, perhaps
by magnetic fields, so that it occurs on a timescale considerably
longer than the dynamical timecsale \citep[e.g][]{1989ApJ...345..782M}
 - recent re-appraisals of the
characteristic lifetime of GMCs suggest that star formation is
rapid but also swiftly terminated \citep{2000ApJ...530..277E,2003ApJ...585..398H}.
In this latter model, only a few per cent of the mass of a cloud
ends up being incorporated into stars prior to its dispersal. Simple
estimates - based on homogeneous molecular clouds - suggest that
dispersal either by photoionisation \citep{1979MNRAS.186...59W,1996AAS...188.1708F} or by OB stellar winds \citep{2002MNRAS.337.1299C} would give
rise to a star formation efficiency of this order.\\
\indent To date, however, numerical feedback studies have been limited to
the case of highly idealised cloud structures, i.e. smooth density profiles
with spherical or axial symmetry \citep{1989A&A...216..207Y,
1990ApJ...349..126F,1996ApJ...469..171G}. Axisymmetric calculations
demonstrate how feedback operates preferentially in directions of
steeply declining density and suggest that the net effect of feedback in
clouds that are realistically inhomogeneous may be very different
from that in smoothed density fields. Although some consideration has
been given to the propagation of radiation in inhomogeneous clouds
\citep[e.g.][]{1993MNRAS.264..145H,1993MNRAS.264..161H,1996ApJ...463..681W,2002A&A...392.1081R,2003MNRAS.342..453H} - particularly with regard to explaining spatial variations in
the abundances of photosensitive species -  such calculations
have not hitherto been combined with hydrodynamical star formation
simulations. The reason for this is that radiation transport is most
readily handled by Eulerian schemes, where it is relatively straightforward
to perform integrals along optical paths. Such schemes are however
challenged by the large density contrasts that develop in realistic
star formation simulations, although Adaptive Mesh Refinement (AMR)
methods are becoming an increasingly powerful tool in this regard
\citep{2003ApJ...587...13T,2003RMxAC..15...92K}.\\
\indent In this paper, we present pilot calculations of photoionisation feedback
in SPH simulations of star cluster formation in realistically inhomogeneous
clouds. (These simulations do not currently incorporate feedback
from stellar winds, which may turn out to be a reasonable omission 
on the timescale of these simulations ($\sim 10^5$ years), since there
is some suggestion that the youngest OB stars have anomalously weak winds; 
Martins et al 2004). Our calculations employ an algorithm for the inclusion
of photoionisation in SPH. Briefly,
this method locates the ionisation front at every hydrodynamical timestep
and adjusts the temperature of the gas behind the front accordingly. 
The ionisation front is located by comparing
the flux of ionising photons emitted in the direction of a given particle 
with the integrated recombination rate between the source and that
particle. In this (Str\"omgren volume) technique,  
the integrated recombination rate is calculated {\it as though} the density
structure between the particle and the source represented the radial
variation of a spherically symmetric density distribution. Evidently
this approach depends on the validity of the on-the-spot approximation
(i.e. on the assumption that diffuse ionising photons resulting from
recombinations to the ground state are absorbed locally). Our method also implicitly assumes that the ionisation structure adjusts instantaneously to any changes in the gas distribution due to, for example, outflows. This assumption is valid provided that the timescale on which the gas distribution evolves is longer than the local recombination timescale, which is always the case in the simulations presented here. 
We note the generic similarity between the present algorithm and
that proposed by \cite{2000MNRAS.315..713K}, inasmuch as
both use the on the spot approximation and utilise the SPH neighbour
lists to construct integrated recombination rates in an efficient manner;
however, whereas Kessel-Deynet and Burkert evaluated the evolution of
the ionisation fraction of each particle using time dependent ionisation
equations, our method - by locating a Str\"omgren volume - assigns to each particle a wholly ionised or wholly neutral state at each
timestep. The method turns out to be simpler and more robust
than that of Kessel-Deynet and Burkert and is justified by the sharpness
of the ionisation fronts in the high density environments we consider. A detailed description of the code and the results of one-dimensional tests will be published in a subsequent paper.\\
\indent In reality, although the boundary of the HII region under such conditions is very sharp, the boundary between hot and cold gas is likely to be less so, since the non-ionising photons from the radiation source would produce a photo-dissociated region (PDR) outside the HII region, whose temperature might reach $1000$ K \citep{1998ApJ...501..192D}. The overall effect of this phenomenon is not easy to predict since, on the one hand, the smaller pressure gradient across the ionisation front will retard the HII region's expansion, but the additional thermal energy deposited over a greater volume of the cluster may increase the total mass involved in outflows. We do not treat PDRs in this work but the combined action of HII regions and PDRs would make an interesting subject of further study.\\
\indent We also do not take into account the likely presence of dust in our HII regions. \cite{1989ApJS...69..831W} found evidence of dust in \textit{all} of the ultracompact HII regions in their survey and estimated that the dust was absorbing between $50$ and $90\%$ of the photon flux emitted by the source in each case. Considerable theoretical work has been done on the effect of dust on HII region evolution. \cite{1998ApJ...501..192D} suggest that there exists an additional initial phase of HII region expansion in which the dust content of the ionised gas is either destroyed or expelled by the star's radiation field. Detailed numerical calculations by \cite{2004ApJ...608..282A} suggest that \cite{1989ApJS...69..831W}'s estimates of the fraction of ionising photons absorbed by dust were too large in most cases. In any case, reduction of the photon flux of a source by $90\%$ results in a reduction in the size of the equilibrium Str\"omgren sphere by only about $50\%$. We therefore expect that inclusion of dust in our calculations would have little effect on our results, although it would probably increase the timescale for expansion of our HII regions somewhat.\\
\indent For these pilot simulations, we plumb our photoionisation algorithm
into the simulations of the formation of a massive star cluster by \cite{2002MNRAS.336..659B}. Specifically, we introduce ionising radiation from a single 
source in the cluster core, $\sim 1$ initial free fall time after the
initiation of the collapse. For our present purposes, the important
feature of the proto-cluster at the time that photoionisation switches
on is that the gas distribution is highly inhomogeneous, with the
cluster core lying at the intersection of several high density filaments
which channel an accretion flow into the core. Although the
Bonnell and Bate calculation is scale free (i.e. can be scaled to any
system of isothermal gas containing the same number of initial Jeans
masses in gas and the same number of  point masses initially), the
introduction of feedback necessarily introduces an absolute mass scale
to the calculation. Our approach here has been to apply two different
scalings to the Bonnell and Bate calculations, so as to
explore the efficiency of feedback in relatively high and low density
environments.\\
\indent In Section 2 we describe briefly the calculations performed by Bonnell and Bate 2002 and the way in which we used these calculations to form the initial conditions for our study of photoionising feedback. In Section 3, we describe a high-resolution and a low-resolution calculation of a protocluster in which the gas densities were high ($10^{5}-10^{9}$ cm$^{-3}$). In Section 4, we describe a single low-resolution calculation of a system with much lower gas densities ($10^{3}-10^{6}$ cm$^{-3}$). In Section 5 we compare the two suites of simulations with simple one-dimensional calculations to assess the effect on the efficiency of feedback of the complex internal structure of our protoclusters and of the gravitational field of the ionising sources within them. In Section 6 we go on to draw conclusions.\\
\section{Initial conditions}
Feedback in all its forms is dominated by massive stars, so the study of the effects of feedback requires an understanding of the formation of high-mass stars.
Although the formation of low-mass stars is moderately well-understood, the
formation of high-mass stars is well known to be problematic. In particular, there has been considerable debate as to whether stars more massive than $\sim10$ M$_{\odot}$ can form by conventional accretion, due to the effect of radiation pressure on dust \citep{1986ApJ...310..207W,1987ApJ...319..850W,2002ApJ...569..846Y,2003MNRAS.338..962E,edgar04}.\\
\indent \cite{1998MNRAS.298...93B} proposed an alternative scenario for massive star formation, namely that massive stars form by the collision and merger of lower-mass stars. \cite{2002MNRAS.336..659B} explored this scenario by performing numerical calculations at two resolutions. The high-resolution calculation was run on the UKAFF facility at Leicester University, while the low-resolution run was conducted on a SUN workstation. Their initial 
conditions were a
cluster containing $1000$ M$_{J}$ of isothermal gas and $1000$ low mass stellar `seeds' whose total mass corresponded to $10$ percent of the gas mass and $\approx9$ per cent of the total mass of the system. They arranged the gas and stars in a spherically-symmetric Gaussian 
density distribution with zero velocity initially. Mergers between two stars were permitted if they approached each other within a distance of $10^{-5}$ times the initial cluster radius.\\
\indent In both their calculations, Bonnell and Bate found that accretion increased the stellar density in the cluster core enough to allow
stellar collisions to occur, contributing in both simulations to the formation of a central object sufficiently massive to have a strong
photoionising flux. In this paper, we take the end results (after about a freefall time) of their simulations as the initial conditions of our  own calculations. At this point, about $7$ per cent of the gas had accreted onto the stars, increasing the mean stellar mass by about $70$ per cent. In addition, the mean stellar density increased by a factor of $10$ and the maximum stellar density (defined by the minimum volume required to contain $10$ stars) rose by a factor of $\simeq10^{5}$. This proved to be sufficient to allow $19$ stellar mergers and contributed to the formation of a very massive star in the cluster core. We emphasise that our choice of this system as a testbed for introducing photoionisation feedback is motivated largely by the fact that it shares several features with observed young clusters. In particular, we wish to explore conditions in which the ionising star is located near the cluster centre and in which the gas distribution is highly anisotropic.\\
\indent We conducted three calculations, designated UH-1, SH-1 and SL-1. In all our calculations, we dispensed with stars and gas at large radii and rescaled Bonnell and Bate's calculations by adjusting the length unit in the code to allow us to study two different regimes of behaviour. In UH-1 and SH-1, we study at two different resolutions virtually identical protoclusters whose gas densities are ($\sim10^{4}$ cm$^{-3}$ - $\sim10^{8}$ cm$^{-3}$). UH-1 is based on Bonnell and Bate's UKAFF run and SH-1 on their SUN run. In both of these calculations, we also rescaled the ambient temperature of the neutral gas to retain the absolute value of the Jeans mass. This resulted in ambient temperatures of $\sim1.4$ K, which are obviously unrealistically low, but we point out that varying the value of the ambient temperature by a factor of a few will have a negligible effect on the dynamical evolution of these calculations, since the behaviour of the protoclusters will be dominated by the thermal pressure of the hot ionised gas and the ram pressure of the accretion flows, both of which are several orders of magnitude greater than the ambient thermal pressure even before rescaling. These calculations are described together in Section 3.\\
\indent In SL-1, also based on Bonnell and Bate's SUN run, we simulate a protocluster with somewhat lower gas densities ($\sim10^{3}$ cm$^{-3}$ - $\sim10^{6}$ cm$^{-3}$). We elected not to rescale the ambient temperature in this calculation, instead leaving it at $10$ K. This calculation is described in Section 4. In all calculations, we used our Str\"omgren volume technique to define the HII region at each hydrodynamic timestep, heating the gas within to $10,000$ K. The properties of our three model clusters are given in Table 1.\\
\begin{center}
\begin{table}
\begin{tabular}{|l|l|l|l|}
\hline
 & UH-1 run & SH-1 run & SL-1 run\\
\hline
Particle number & 5$\times10^{5}$ & 5$\times10^{4}$ & 5$\times10^{4}$\\
Total mass (M$_{\odot}$) & 742 & 745 & 745\\
Initial stellar mass (M$_{\odot}$) & 225 & 223 & 223\\
Initial gas mass (M$_{\odot}$) & 517 & 522 & 522\\
Initial radius (pc) & 1.4 & 1.4 & 2.8\\
Initial Jeans mass (M$_{\odot}$)& 0.8 & 0.8 & 40.5\\
Core gas density (cm$^{-3}$) & $\sim10^{9}$ & $\sim10^{8}$ & $\sim10^{6}$\\
Mean gas density (cm$^{-3}$) & $\sim10^{4}$ & $\sim10^{4}$ & $\sim10^{3}$\\
Freefall time (yr) & $1\times10^{6}$ & $1\times10^{6}$ & $2\times10^{6}$\\
Source luminosity (s$^{-1}$) & $6\times10^{49}$ & $6\times10^{49}$ & $2\times10^{49}$\\
\hline
\end{tabular}
\caption{Parameters of our runs. All gas densities quoted are atomic hydrogen number densities.}
\end{table}
\end{center}
\section{Simulations of high-density initial conditions}
We describe the UH-1 and SH-1 runs together in this section since they are essentially simulations of the same protocluster differing only in their numerical resolution. In common with \cite{2002MNRAS.336..659B}, we conducted our high-resolution UH-1 simulation on the UK Astrophysical Fluids Facility's SGI Origin 3800 and the corresponding low-resolution one on a SUN workstation. The initial conditions for the UH-1 and SH-1 simulations, given in Table 1, were quantitatively almost identical. The ionising source in these calculations has an initial mass of $\sim30$ M$_{\odot}$. We regard it as a point source of ionising photons and 
assign to it an ionising flux of $6\times10^{49}$ s$^{-1}$. This is a rather generous output
even for such an O-star ($2\times 10^{49}$ s$^{-1}$ would be more realistic), but is justified, since the ionising star in these simulations grows considerably in mass (to a few$\times100M_{\odot}$) through further accretion (see \cite{1998ApJ...501..192D} for a table of O-star ionising luminosities). We note that this luminosity is close to the source's classical Eddington luminosity, but accretion continues along the gaseous filaments, since the inhomogenous structure of the gas increases the source's effective Eddington luminosity (see \cite{2001MNRAS.326..126S} for a discussion of super-Eddington accretion in inhomogenous environments).\\
\subsection{The UH-1 run}
\begin{figure}
\includegraphics[width=0.45\textwidth]{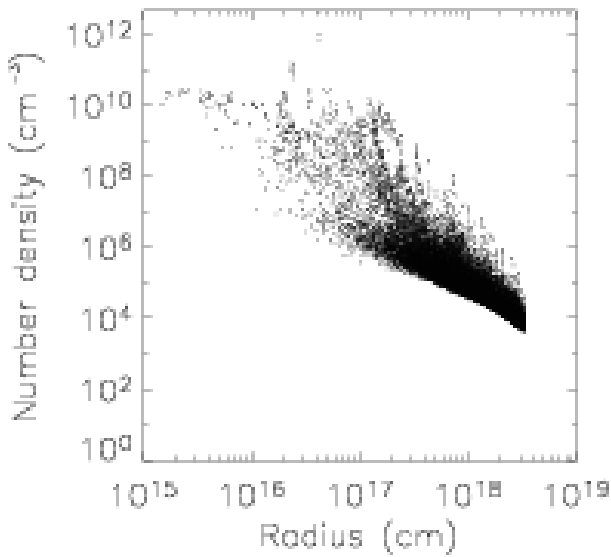}
\caption{Initial radial density profile of the UH-1 calculation. Only every tenth particle has been plotted.}
\end{figure}
\indent In Fig. 1, we plot the initial radial density profile of the UH-1 protocluster. There is clearly an enormous range of gas 
densities in the system. The profile
approximately follows a power-law with slope $\alpha\approx-2$, but there is
evidently a great deal of substructure superposed on this general configuration.
This can be seen in detail in Fig. 2 which shows a column-density
map of the protocluster (the ionising source is at the centre of the image).\\
\begin{figure}
\includegraphics[width=0.45\textwidth]{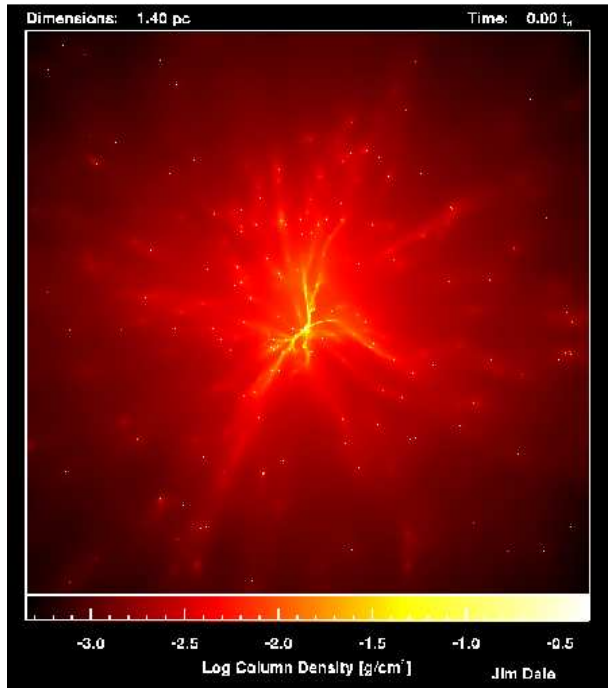}
\caption{Column density map of the inner $1.4$ pc$\times1.4$ pc of the initial 
conditions for our UH-1 calculation as viewed along the $z$-axis.} 
\end{figure}
\indent It can be seen that the initially-smooth gas in Bonnell and Bate's original calculation has developed a complex filamentary structure which approximately radiates outwards from the core. The distribution of residual gas closely follows
that of the stars, with stars being preferentially located inside gaseous
filaments. These structures are due to gravitational 
instabilities in the gas which amplify pre-existing shot noise in the stellar seed
distribution. One of the aims of this study is to see what effect this structure has on the efficacy of feedback.\\
\indent  The gas is distributed highly anisotropically with respect to the O-star at the centre of the protocluster. Figure 3 shows a $\theta,\phi$ column density 
map of the cloud as seen from the ionising source. One expects that this structure will result in the distance to which ionising radiation can penetrate the protocluster to be similarly anisotropic and our simulations demonstrate that this is true.\\
\begin{figure}
\includegraphics[width=0.45\textwidth]{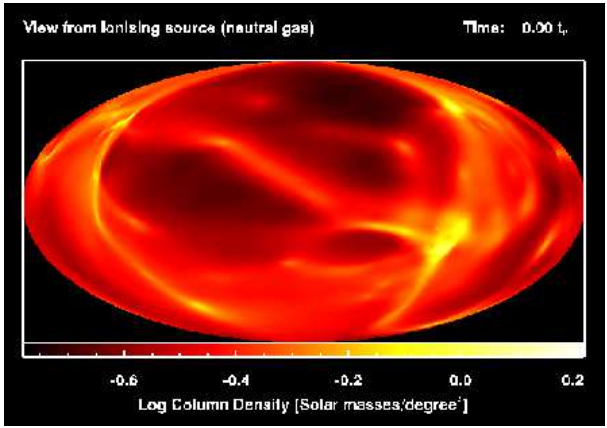}
\caption{$\theta, \phi$ column density map of the gas distribution in the
UH-1 cluster at the beginning of the simulation as seen from the ionising source. Only material within $0.2$ pc has been included in this map in order to emphasise the anisotropic conditions encountered by the source's radiation field.}
\end{figure}
\indent We found that ionising radiation is able to escape from
the immediate vicinity of the massive star in only a few directions, forming
several outflows and corresponding elongated HII regions. The
radiation and thus the outflows are loosely collimated by the dense gas near the
star. By restricting the directions in which radiation can escape from the
source, a small mass of gas distributed very close to the ionising source is able to
dominate the evolution of the ionised region and thus to control its effect on
the rest of the cloud. Bonnell and Bate suggested such collimation as a possible 
explanation for the poorly-collimated outflows observed around young massive 
stars. This work lends considerable support to this proposal. However, we note here that this collimation may be due in part to noise in the particle distribution close to the source, since a single particle at a small radius is potentially able to shield a large number of particles further out in the cloud. This importance of this effect could be studied by splitting gas particles close to the source into $\sim10$ (for example) smaller, lower-mass particles, but the presence of particles of very different masses in the same region of space can give unphysical results in SPH. We have not investigated this issue.\\
\indent The end result of our calculation is shown in Fig. 4 at which point the ionising source has been radiating for $\sim1.7\times10^{5}$ yr, 
$\sim17\%$ of the cluster's global freefall time.\\
\begin{figure}
\includegraphics[width=0.50\textwidth]{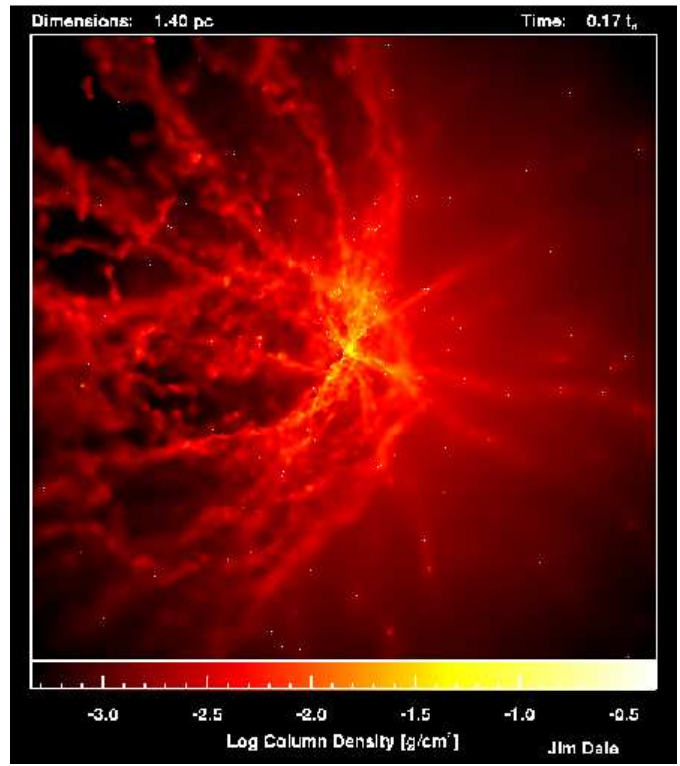}
\caption{Column density map showing the end result of the UH-1
calculation after $0.17$ t$_{ff}$, as viewed along the $z$-axis. Note the
asymmetric distribution of the outflows.}
\end{figure}
\subsubsection{Outflows}
The outflow structure is evidently very complex. Watching an animation of the
simulation allows six or seven individual radial flows to be identified. These
outflows are visible as the dark regions in Figure 5, 
a second $\theta,\phi$ plot of the gas column density seen from the source, 
generated at the same epoch as Fig. 4. The disposition of the outflows is 
interesting in that they are all confined to the left-hand half of the
cloud. Hence, the left hemisphere of the cloud has a radically different structure by the end of the
calculation, whereas the right hemisphere is almost untouched.\\
\begin{figure}
\includegraphics[width=0.45\textwidth]{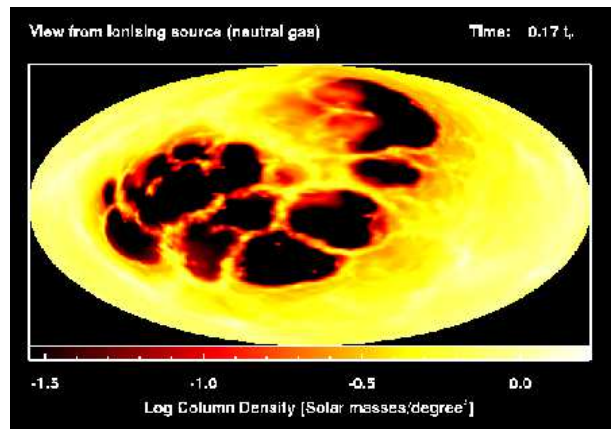}
\caption{A $\theta,\phi$ column density map of the neutral gas distribution as seen from
the source in the UH-1 calculation at $t=0.17t_{ff}$. All neutral gas has been included in this map.}
\end{figure}
\indent The outflows are weakly collimated. Their opening angles can be
estimated by eye from Figure 5 and are
approximately $10^{\circ}-20^{\circ}$. The ionised gas in the outflows expands azimuthally as well as radially, sweeping up neutral material and generating numerous
fresh filamentary structures, particularly where neutral gas is caught between two expanding bubbles. These filaments are also visible in Fig. 5 as the narrow bright features within the dark 
areas of the plot. The filamentary structures subsequently fragment into
strings of clumps whose densities increase with time, as they are immersed in hot ionised gas and are
therefore under
strong compression. At the point where we
were forced to stop the simulation, the entire left-hand hemisphere of the
protocluster is a chaotic assembly of chains of 
isolated clumps and filaments still in the process of subfragmentation. 
The right-hand hemisphere of the cloud remains largely 
untouched because of shielding by very dense material just to the right of the ionising source. It can be seen in the animation that the source is beginning to clear away this material and it is likely that the protocluster would become more homogeneous if 
it were possible to continue the simulation. However, we were forced to terminate the simulation prematurely because numerous particles acquired timesteps which were prohibitively short, slowing the calculation dramatically. This is due to the development of strong density enhancements owing to the increased fragmentation.\\
\indent We compare the mass entrained in the outflows and
the momentum this mass carries with outflows observed in real
star-forming regions. In Fig. 6, we plot histograms showing the masses of
gas involved in inflows and outflows at the beginning and at the end of the
simulation. Initially, most of the gas in the system is infalling at speeds of $5$ kms$^{-1}$ or less. The outflows are clearly visible in the overlaid later plot as material 
with high positive radial velocities. The total mass involved
in the outflows is $\sim30$M$_{\odot}$, about seven percent of the total gas mass.
Only about one third of the gas involved in the outflows is ionised, the
remainder being swept-up neutral gas. The mean 
rate at which the outflows are removing mass from the cluster core is
$\approx2\times10^{-4}$ M$_{\odot}$ yr$^{-1}$. The total outflow momentum is $\approx 10^{3}$ M$_{\odot}$ kms$^{-1}$, in good agreement with observed values as tabulated by \cite{1997ApJ...479L..59C} for stars in the spectral range B0-O4.\\
\indent Although this agreement with the parameters of observed outflows is encouraging, we note here that the outflows in our simulations, which are driven by anisotropic heating, are not in general bipolar in nature, since their number and symmetry is determined purely by details of the initial gas structure near the radiation source. At first sight, this would appear to disfavour such a mechanism since outflow sources from young massive stars are usually described as being bipolar \citep[e.g.][]{2002ApJ...576..313K,2002A&A...387..931B}. However, as noted by \cite{1997ApJ...482..355S}, this designation can be misleading, since outflow lobes may not be symmetric and some outflows appear to be monopolar. The outflows described here clearly cannot be compared to the highly-collimated sources recently discovered in pre-UCHII regions (\cite{2002A&A...387..931B}, \cite{2004A&A...425..981D}) but they may be compatible with some of the less collimated objects described by \cite{1997ApJ...482..355S}. It is currently unclear what fraction of observed outflows could be explained by the mechanism we discuss here, and what fraction require the conventional scenario of bipolar flows emanating from an accretion disc. 
\begin{figure}
\includegraphics[width=0.45\textwidth]{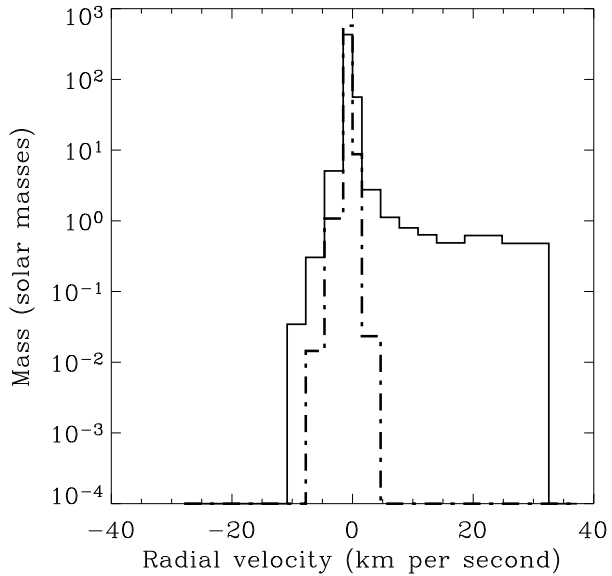}
\caption{Histograms showing the initial distribution of velocities with mass
(dotted lines) and the final distribution including the
outflows driven by the thermal pressure of the ionised gas in the UH-1 calculation, as described in the text (solid lines).}
\end{figure}
\subsubsection{HII regions}
\indent The structure and behaviour of the ionised
gas in this simulation is unusual for two reasons. Firstly, the HII region 
possesses a multi-lobed structure due to the beaming effects mentioned in the preceding
paragraphs. An example is shown in Figure 7, showing clearly that the HII region(s) generated in these calculations bear little resemblance to the classical picture of the Str\"omgren sphere. Although unusual, the HII region depicted in Figure 7 is in appearance not unlike some of the `cometary' ultracompact HII regions described in Wood and Churchwell's seminal paper on the subject \citep{1989ApJS...69..831W}, in particular G29.96-0.02 (shown on page 863 of \cite{1989ApJS...69..831W}) and G43.89-0.78 (shown on page 867).\\
\indent The second unusual property of our HII regions is their variability. In the early stages of the simulation, the 
HII regions are observed to `flicker' on short timescales (a few hundred to a few thousand years), 
frequently disappearing completely, only to regrow a short time later. This can
be seen in Fig. 8 where we plot the fraction of the protocluster's gas
which is ionised as a function of time. Although the general tendency is for the ionisation fraction to increase, it exhibits numerous sharp rises and falls, particularly in the early stages of the calculation.\\
\indent Figure 8 implies, particularly in the early stages of the simulation, that the HII region contains small numbers of SPH particles, sometimes below the canonical resolution limit of $\sim50$, corresponding to a particle and its neighbours. We conducted convergence tests in a uniform spherically symmetric model cloud to determine what effect such apparent under-resolution might have on the expansion of the HII regions in our simulations. We found that we were able to reproduce the well-known Spitzer solution for the expansion of a spherical HII region in a uniform cloud (\cite{1978ppim.book.....S}) even when the HII region initially contained as few as three (!) SPH particles. We concluded from this that the resolution requirements pertinent to modelling the expansion of HII regions are rather less stringent than those relevant to other physical processes, although clearly the accuracy with which the detailed geometry of HII regions can be modelled is dependent on the available resolution. 

\begin{figure}
\includegraphics[width=0.50\textwidth]{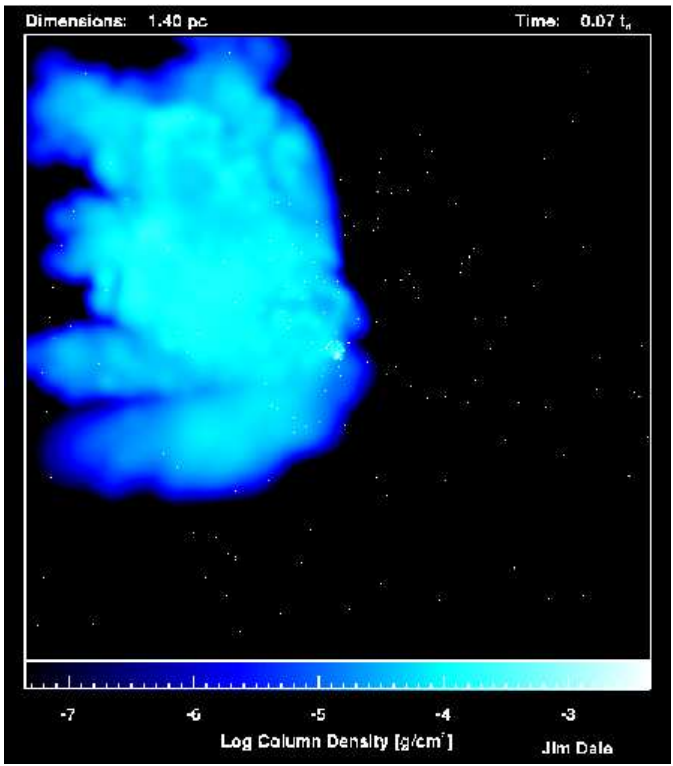}
\caption{Column density map of the ionised gas in the UH-1 simulation $0.07$ t$_{ff}$ after source ignition revealing the unusual shape of the HII region.}
\end{figure}

\begin{figure}
\includegraphics[width=0.45\textwidth]{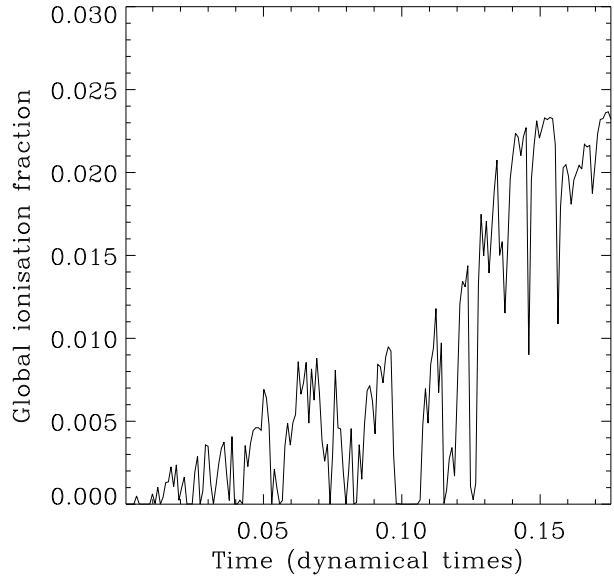}
\caption{A plot of the UH-1 protocluster's global ionisation fraction as a function
of time, revealing the flickering process described in the text. The dynamical time of the system is $10^{6}$ yr.}
\end{figure}
\indent The cause of the flickering behaviour is the 
rapid and non-uniform accretion taking place in the core of the 
protocluster, mostly along the dense filaments. Although, in the simulations, this effect is partly shaped by finite resolution near  the source (i.e. the effect of individual SPH particles), it is more generally the case that small fluctuations in the amount of neutral matter near the source can have a profound effect on the propagation of ionising radiation to large radii. At early times, the interplay of the expanding HII regions and the accretion flows results in the source region being intermittently inundated by inflowing neutral gas on a characteristic timescale similar to the freefall timescale in the cluster core. Since this timescale is long compared with the recombination timescale anywhere in the cluster, the HII regions are able to respond by switching on and off on this timescale (see Figure 8). However, after about $10^{5}$ yr, a quasi-steady flow pattern is set up, in which the neutral flow never completely swamps the core region and the mass of ionised gas thereafter grows steadily with time.\\
\indent Once the HII region has reached its stable phase, it is able to have a dramatic effect on the structure of the gas in the protocluster, since ionised gas is able to
permeate a significant fraction of the cluster's volume (although the ionisation
fraction in this simulation was never more than three
percent). We observed the emergence of a great deal of novel structure, but we were not able to follow the protocluster's
evolution long enough to study the gross dynamical effects of photoionising
feedback. We therefore turn to the low-resolution SH-1 calculation which was able to
run for somewhat longer.\\
\subsection{The SH-1 run}
\indent The SH-1 calculation was started from similar 
(although not identical) initial conditions to those of the UKAFF run and was able to continue for considerably longer, allowing us to study the long-term dynamical impact of the HII region on our
protocluster and to determine the effect of the fragmentation process observed
in the high-resolution run on the star formation within the cloud. The initial
conditions for this calculations are shown in Fig. 9. They are clearly
qualitatively similar to those of the high-resolution run, as can be seen by
comparing Figures 2 and 9.\\
\begin{figure}
\includegraphics[width=0.45\textwidth]{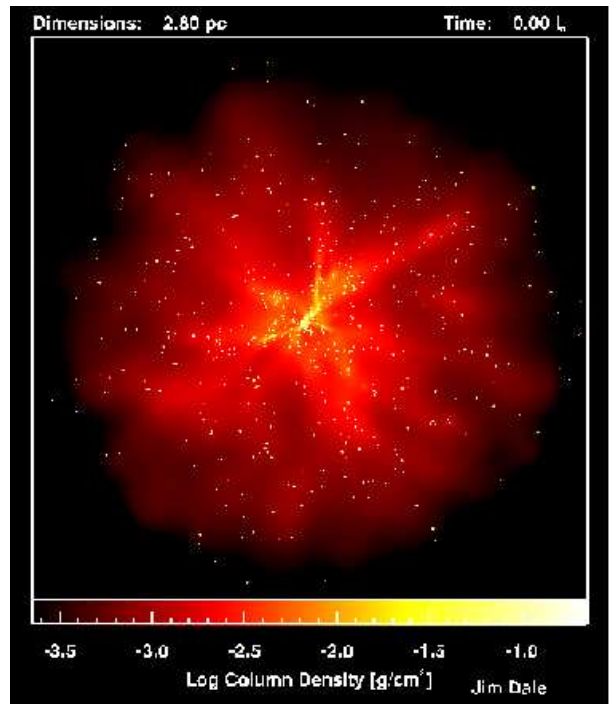}
\caption{Column density map of the initial conditions for our SH-1
calculation as viewed along the $z$-axis.}
\end{figure}
\indent The evolution of the low-resolution simulation was similar to the
UKAFF calculation, as shown in Figure 11. We again observed several weakly-collimated outflows driving
bubbles of ionised gas into the protocluster's ambient neutral gas, generating
filamentary and beadlike structures as they went. We also observed the
phenomenon of HII-region flickering, depicted in Figure 10, again dying away as the
ionising source was able to expel some of the dense gas from its immediate vicinity and
deflect the accretion flows away from the core. However, in this case, we are
more interested in the later evolution of the system.\\

\subsubsection{Accretion, energy uptake and cluster survival}
\begin{figure}
\includegraphics[width=0.45\textwidth]{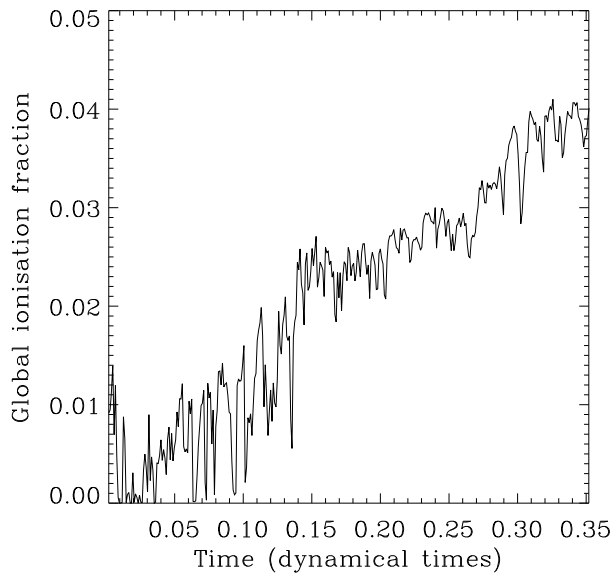}
\caption{A plot of the global ionisation fraction as a function of time for our SH-1 calculation. As in the UH-1 run, we observe that the HII region initially flickers rapidly before settling down to a stable expansion phase. The dynamical time of the system is $10^{6}$ yr.}
\end{figure}
\begin{figure}
\includegraphics[width=0.45\textwidth]{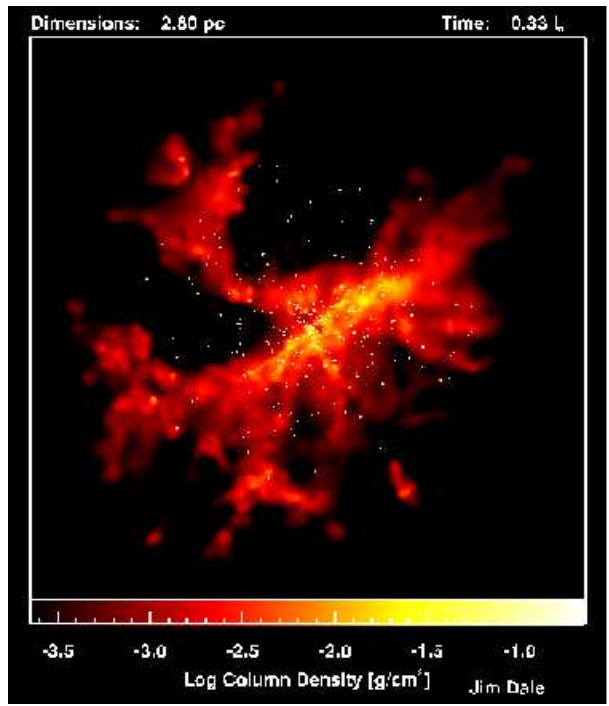}
\caption{Column density map of the end result our SH-1
calculation after $0.33$ t$_{ff}$, as viewed along the $z$-axis.}
\end{figure}
\indent The most obvious question to ask is whether the radiation from our
massive star is able to prevent the cluster from becoming permanently gravitationally bound.
At the beginning of our simulations, the cluster is bound, but the ratio of gas
mass to stellar mass is $\sim2.5:1$, so rapid expulsion of significant
quantities of gas could unbind the cluster. Since accretion is still in progress in the core of the cluster, this ratio is clearly 
approaching $1:1$ as time progresses. If the ratio of gas mass to stellar 
mass reaches $1:1$ while the system is still globally bound, the
cluster, or at least its stellar content, will remain bound regardless of what
happens to the remaining gas. The question of whether feedback can
unbind the cluster therefore reduces to seeing if the accretion flows can be 
stopped before this occurs. In Fig. 12, we show how the total masses of the stellar
and gaseous components of the protocluster evolve with time in the SH-1 calculation. We compare the SH-1
calculation with another one identical except that the ionising source was left
switched off. In both plots, all the gas and all the stars in the clusters are included whether still bound to the cluster or not, so the change in the masses of the two components is due purely to accretion.\\
\begin{figure}
\includegraphics[width=0.45\textwidth]{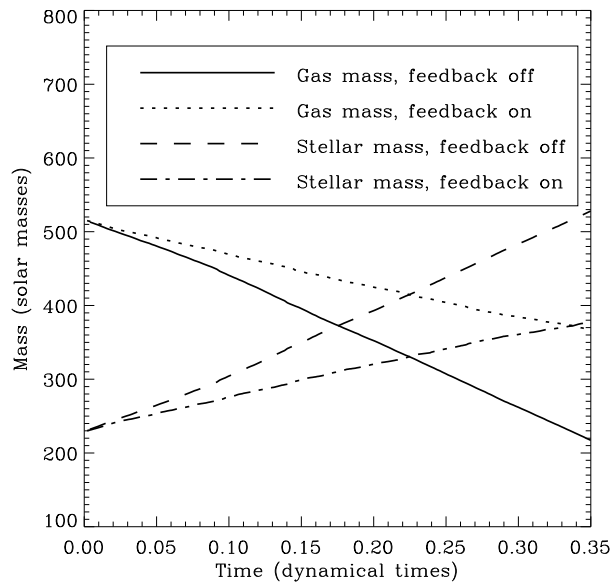}
\caption{Comparison of the evolution of the masses of the SH-1 cluster's stellar and
gaseous components in runs with feedback active and inactive. The dynamical time of the system is $10^{6}$ yr.}
\end{figure}
\indent The global accretion rates in the two simulations are given by the
gradients of the lines in Fig. 12. The accretion rate in the case with
feedback active is lower than in the run without feedback but accretion is never 
halted because the thermal pressure of the
ionised gas is considerably less than the ram pressure of the accretion flows. 
The stellar and gas masses reach equality at a 
later time with feedback active but examination of the energies of the cluster's stellar and gaseous components revealed
that all the stars and $\sim 83\%$ of the gas are
still bound at this point in time, so photoionising
feedback has failed to unbind the cluster.\\
\indent We now compare the stellar energy input with the thermal and
kinetic energies of the gas to see how efficiently the energy emitted by the ionising source is being taken up by the system and we compare the total energy uptake with the cluster's gravitational self-energy. 
The gravitational potential energy of an object of mass $M$ and radius $r$ is
given approximately by
\begin{eqnarray}
E_{grav}\simeq\frac{GM^{2}}{r}
\end{eqnarray}
Our protocluster has a total mass of $740$ M$_{\odot}$ and a radius of 
$4.2\times10^{18}$ cm, giving $E_{grav}\simeq3\times10^{46}$ erg.
In Fig. 13, we plot the gas thermal and
kinetic energies against time along with the energy input by the star since its
ignition (taking this to be the energy contained in $6\times10^{49}$ $13.6$ 
eV photons s$^{-1}$ $\times$ (time since ignition)). Two things are immediately
apparent form Fig. 13:
firstly, the photoionisation process in this cluster is extremely inefficient, with
$<0.1\%$ of the energy injected by the star being retained. We have assumed that all recombinations of ions with electrons that initially deposit the electron in an atomic energy state other than the ground state result in the emission of a photon or a series of photons to which the gas is optically thin. Such recombinations therefore constitute an energy sink. In addition, at the end of the calculation, $\sim20\%$ of the ionising flux is escaping from the cloud directly, through holes in the gas distribution.\\
\indent Secondly, despite the inefficient coupling, the star apparently \emph{has} 
deposited enough energy into the gas to unbind the cluster as a whole.\\
\begin{figure}
\includegraphics[width=0.45\textwidth]{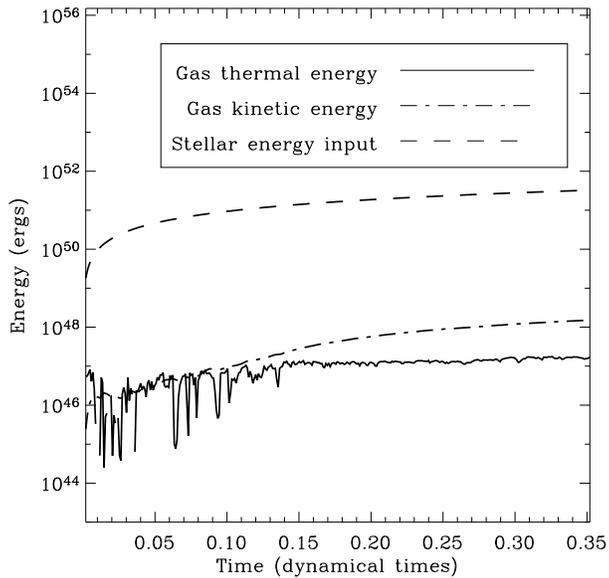}
\caption{Comparison of the stellar energy input with the gas thermal and kinetic
energies in the SH-1 calculation. The dynamical time of the system is $10^{6}$ yr.}
\end{figure}
\indent This leads to the obvious question of why the cluster does not become
globally unbound. The answer is simply that, although sufficient energy has been
absorbed by the cluster gas to exceed the cluster's total gravitational
binding energy, this
energy is distributed throughout a small proportion ($\sim17\%$) of the
gas. As can be seen in Fig. 14, the ionised gas is escaping from the
cluster, entraining some neutral gas with it, and carrying thermal and kinetic
energy away, so this energy has little effect on the cloud as 
a whole. Although the energetic material becomes unbound, the remainder,
constituting over $83\%$ of the cluster's gaseous component, is
still firmly bound within the cluster's gravitational potential well.\\
\begin{figure}
\includegraphics[width=0.45\textwidth]{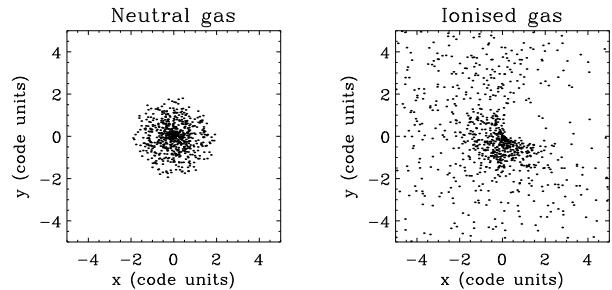}
\caption{xy-plots of the positions of neutral and ionised gas particles in the SH-1 run,
revealing that the hot ionised gas is able to effectively escape the 
protocluster.}
\end{figure}
\indent Our neglect of the likely existence of a PDR outside the HII region in these calculations clearly affects the energetics of our cluster. The presence of a PDR would result in the thermal energy of the cluster gas being higher and may also increase the kinetic energy of the gas by involving a larger fraction of it in outflows. However, the results presented in this section suggest that \textit{where} in the cluster thermal energy is deposited is at least as important as the bare quantity of energy. The expansion of the HII region in this protocluster was controlled by the dense material and accretion flows near the radiation source and it is likely that the PDR would be similarly constrained. We think it unlikely that modelling the PDR would substantially alter our conclusions.\\
\subsubsection{Star formation}
\indent In none of our simulations was the mass resolution good enough to 
follow the star formation process in detail, but we can draw qualitative 
conclusions about the effect of feedback from the evolution of the cluster's
mean Jeans mass, depicted in Figure 15, where 
\begin{figure}
\includegraphics[width=0.45\textwidth]{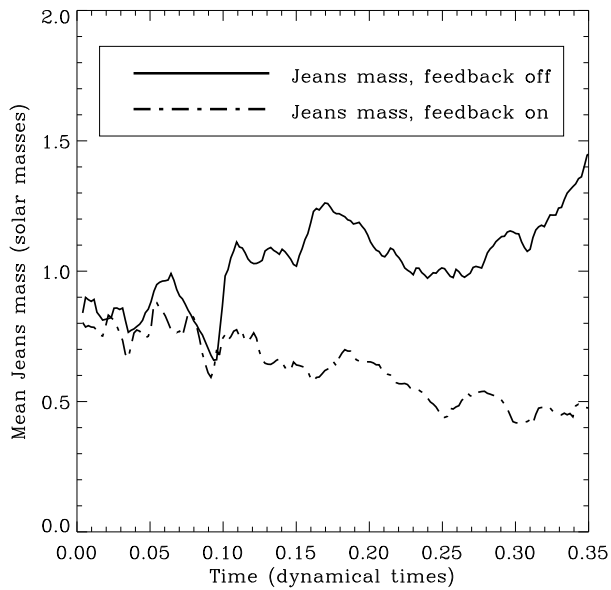}
\caption{Evolution of the mean Jeans mass within the SH-1 cluster. The dynamical time of the system is $10^{6}$ yr.}
\end{figure}
we have again made a comparison with the control simulation without feedback. In both cases, the mean Jeans mass was estimated using the mean density of the \textit{neutral} component of the cluster gas only, comprising not less than $95$ per cent of the total gas mass of the feedback (and $100$ per cent of that in the control run), taking the neutral gas in both runs to have the initial ambient temperature. The evolution of the mean Jeans mass in the two systems is initially similar, but becomes very different after $\sim0.1$ $t_{ff}$. Although both plots exhibit rapid fluctuations, the general trend in the control run is for the Jeans mass to increase, whereas in the feedback run, the Jeans mass decreases. The former result is somewhat counterintuitive: since the system is collapsing unimpeded, one might expect the mean gas density to rise and the Jeans mass to fall (given that the temperature is constant). However, because of the rapid accretion taking place in the control simulation and because high-density gas in the cluster core is accreted more rapidly than low-density gas further out in the cluster, the mean gas density in this calculation actually \textit{decreases} overall.\\
\indent In the feedback calculation, the accretion rate is considerably lower, as shown in Figure 12 and it is in the core that the accretion rate is most affected. In addition, the spread of hot ionised gas throughout much of the cluster's volume results in the compression and fragmentation of much of the system's remaining neutral gas - this process can be clearly seen in Figure 4 by comparing the highly-fragmented left-hand-side of the cloud with the almost-untouched right-hand side. The net result of these effects is that the mean gas density in the feedback calculation \textit{increases}, driving the Jeans mass down (non-monotonically) from its initial value of $\approx0.8$ M$_{\odot}$ to 
$\approx0.4$ M$_{\odot}$. Although the mass resolution of the calculation ($\sim0.5$ M$_{\odot}$) is not sufficient to follow the fragmention/star-formation process (and indeed only a few new stars were observed to form during the simulation), we can infer from this plot that the overall effect of the ionising source is 
to decrease the protocluster's mean Jeans mass and thus to promote fragmentation. We cannot, however, make any statements regarding how many new stars will actually form.\\
\section{Simulations of low-density initial conditions}
\indent In the previous section, we studied a cluster in which the feedback
effects of the massive stars were strongly constrained by the physical structure
and dynamical properties of the gaseous component of the system. We now consider a protocluster in which the gas density is approximately an order of magnitude lower and on which we therefore expect feedback to have a greater impact.\\
\subsection{The SL-1 run}
\indent In our SL-1
simulation, we study a cluster in which the gas density is considerably lower (as we have increased the physical size of the cluster), in an
attempt to give feedback an opportunity to have a greater influence
on the protocluster's dynamical evolution. The ionising source in this simulation had an initial mass of $\sim30$ M$_{\odot}$, rising to $\sim60$ M$_{\odot}$ by the end of the run. It was assigned an ionising photon flux of $2\times10^{49}$ s$^{-1}$.\\
\begin{figure}
\includegraphics[width=0.50\textwidth]{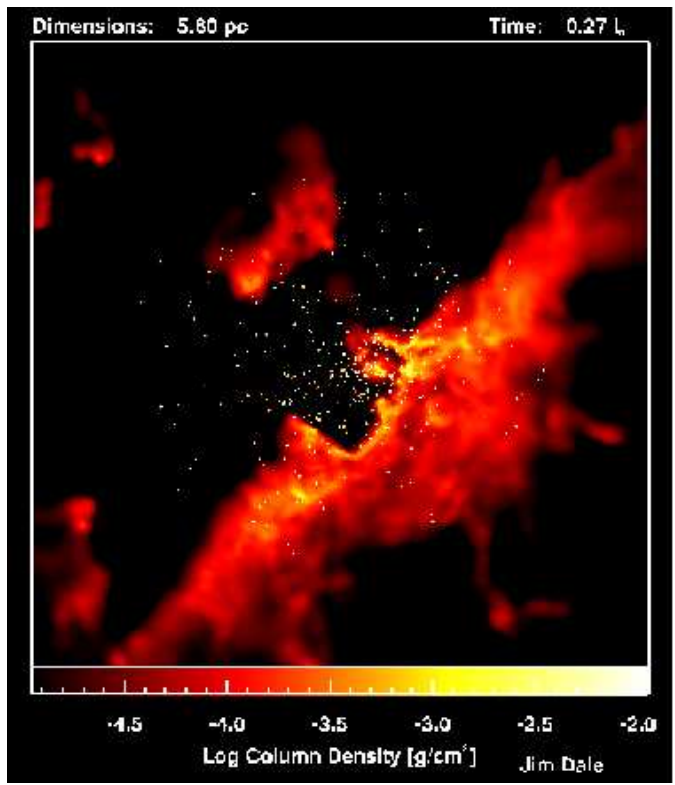}
\caption{Column density map of the end result of the SL-1 run after 0.27$t_{ff}$, as viewed along the $z$-axis.}
\end{figure}
\subsubsection{Accretion, energy uptake and cluster survival}
\indent The evolution of this simulation is markedly different from that of either of the foregoing calculations. Fig. 16 shows a column-density map of the end result. Although superficially similar to Fig. 11, careful
inspection shows that photoionisation has been more successful 
in expelling gas from the core of the protocluster in the low density 
simulation. As can be seen by close examination of Figs. 11 and 
16, there is a higher degree of segregation between the stars and the gas in
the SL-1 simulation and this has been achieved on a shorter timescale than in the SH-1 calculation. Because of the lower gas densities in the SL-1 system, the radiation source is able to partially ionise and thus disrupt the radial accretion flows. Fig. 17 shows a plot of the global ionisation fraction in the SL-1 run. Although the flickering phenomenon observed in the UH-1 and SH-1 calculations does occur in this run, it dies away much more rapidly, allowing the ionisation fraction to rise more monotonically and to attain higher values, although only by a factor of a few. The disruption of the accretion flows has the natural consequence that the global accretion
rate in the SL-1 run is lower. Figure 18 shows that feedback quickly brings accretion to an almost complete halt, leaving the stellar:gas mass ratio 
little changed from its initial value.\\
\begin{figure}
\includegraphics[width=0.45\textwidth]{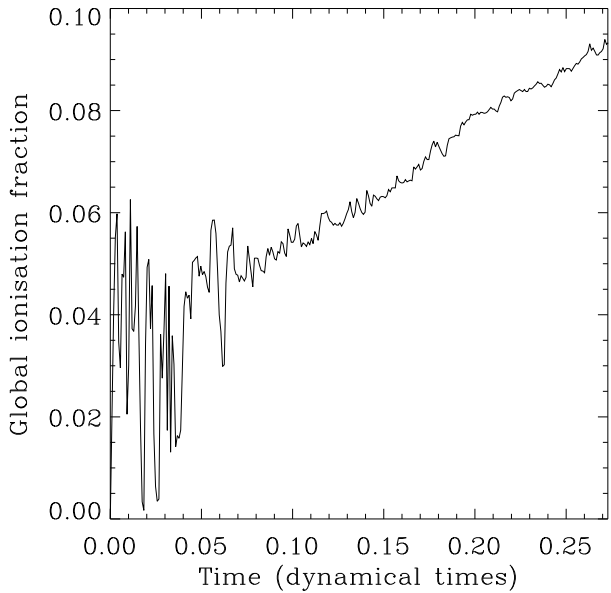}
\caption{Evolution of the global ionisation fraction in the SL-1 calculation. The dynamical time of the system is $2\times10^{6}$ yr.}
\end{figure}
\begin{figure}
\includegraphics[width=0.45\textwidth]{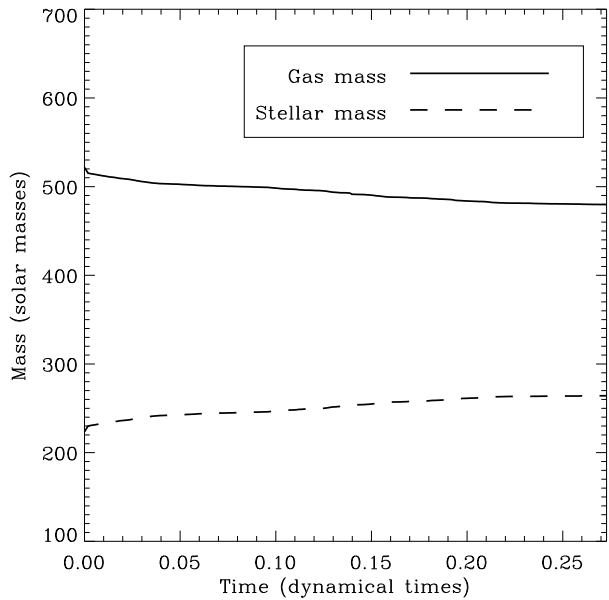}
\caption{The accretion behaviour of the SL-1 calculation. The dynamical time of the system is $2\times10^{6}$ yr.}
\end{figure}
\begin{figure}
\includegraphics[width=0.45\textwidth]{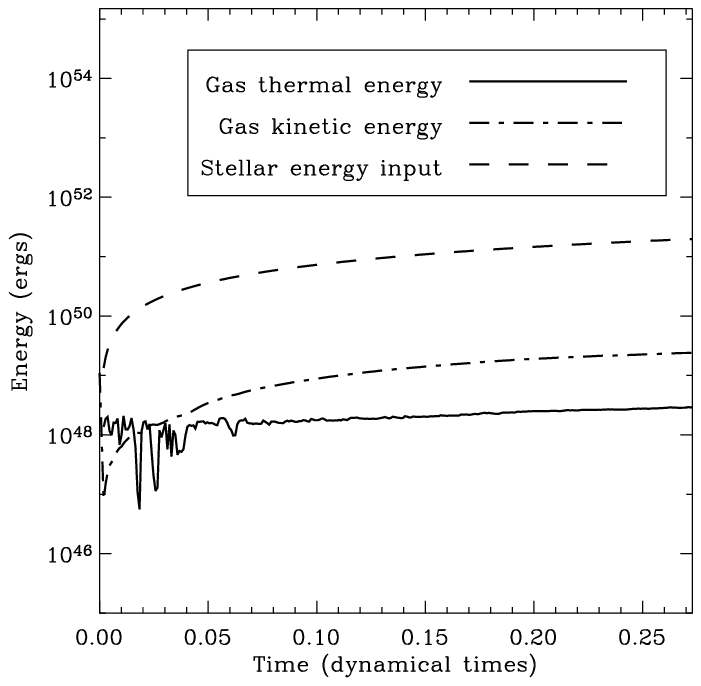}
\caption{Comparison of the stellar energy input with the gas thermal and kinetic energies in the SL-1 run. The dynamical time of the system is $2\times10^{6}$ yr.}
\end{figure}
\indent As shown in Figure 19, the energy transfer in this calculation is somewhat more efficient than in the SH-1 simulation, despite the fact that $\sim70\%$ of the ionising flux is escaping from the system at the end of the run. In addition, the gravitational binding energy of this system is somewhat lower ($\sim10^{46}$ erg), so the SL-1 protocluster is `easier' to unbind. This is reflected in the dynamical end-states of the calculations. In the SH-1 calculation, the fraction of gas still gravitationally bound at the `point of no return' where the total stellar mass comes to exceed the total gas mass was $83\%$ and accretion was still in progress, so the SH-1 protocluster was almost certain to remain globally bound. By contrast in the SL-1 protocluster, accretion is almost halted by feedback and the bound gas fraction drops to less than
$65\%$ by the end of the simulation, $\sim0.27$ freefall times after source ignition. $5\%$ of the
stars are also unbound at this point.  We expect the fraction of unbound stars to 
rise as more gas is expelled from the system, but it is not possible to say on 
the basis of this calculation whether any of the cluster's stellar population
will remain bound.\\
\indent As in the SH-1 simulation, the energetics of this system have been modelled ignoring the likely existence of a PDR surrounding the HII region which would result in the deposition of a larger quantity of thermal energy in the gas and over a larger volume. It is probable that modelling this effect would lead the SL-1 protocluster to become unbound faster but, again, we do not feel that our conclusions would be significantly different.
\subsubsection{Star formation}
\begin{figure}
\includegraphics[width=0.45\textwidth]{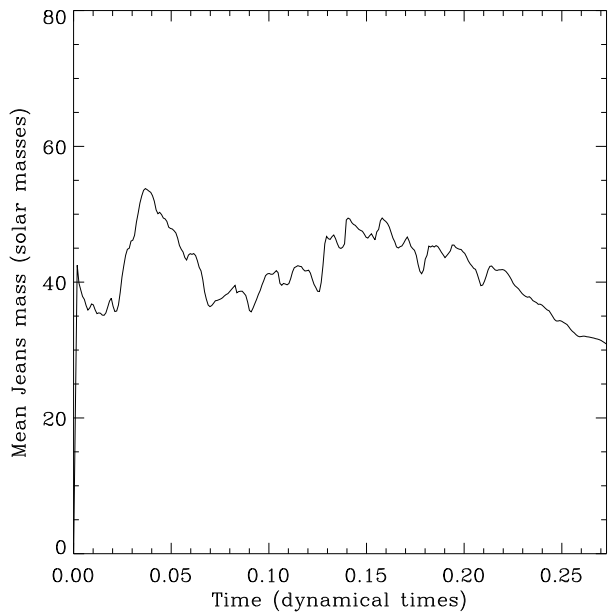}
\caption{The evolution of the mean Jeans mass in the SL-1 calculation.}
\end{figure}
We qualitatively assess the impact of feedback on star formation in the SL-1 cluster by examining the evolution of the mean Jeans mass, again estimated using the mean density of the protocluster's \textit{neutral} gas content only and taking its temperature to be the initial ambient temperature. The result is shown in Figure 20. After approximately $0.05$ t$_{ff}$, the mean Jeans mass increases sharply by about fifty percent. This is due destruction of dense filaments of neutral gas near the radiation source by expanding ionised gas. Although this process continues throughout the calculation, it is counterbalanced by the compression of neutral gas further out in the protocluster as it is swept up by the growing HII region. Comparison of the SH-1 and SL-1 calculations reveals that feedback has two competing effects on the gas in protoclusters. Which effect dominates is determined by the gas density of the system. Feedback can decrease the gas density in the cluster core by destroying the structure there, but this effect was not observed in the high-density calculations because this material was simply too dense. Conversely, feedback can raise the gas density in the outer reaches of the cluster by sweeping up neutral gas and generating new structure. This latter effect dominated in UH-1 and SH-1 calculations, driving the global mean gas density up and the mean Jeans mass down. In the SL-1 calculation, the two effects roughly cancel each other for most of the run and the mean Jeans mass changes little. However, after $0.2$ t$_{ff}$, disruption of the dense structure in the SL-1 cluster core is complete and compression of material further out in the cluster begins to dominate. The impact of feedback on the star-formation process in this system is less clear-cut than in the previous simulations but, in the later stages of the calculation, feedback is again aiding the fragmentation process. Although the Jeans mass is well-resolved in this calculation, we did not observe the formation of any new stars. This is due to the fact that a Jeans mass of $\sim40$ M$_{\odot}$ represents a significant fraction of the cluster's total residual gas content and it is very difficult to gather such a large quantity of material into one place.\\
\indent The competition between the constructive and destructive impacts of feedback seen in these simulations is also apparent in the behaviour of systems such as M16 and NGC 3603 in which the effects of massive stars on molecular clouds can be observed directly. Based on their observations of NGC 3603, \cite{2001RMxAA..37...39T} conclude that, while the cluster HD 97950 is evidently destroying its natal molecular cloud, the existence of embedded infrared objects near the ionisation fronts also implies that triggered star formation is in progress in NGC 3603. \cite{1996AJ....111.2349H} observe that the erosion of the famous gas pillars in the Eagle nebula is `fixing' the masses of the stars already formed by preventing them from accreting additional material, just as occurred in our SL-1 calculation. However, they also suggest that radiatively-driven implosion may be responsible for stimulating the formation of the young stars in M16. However, \cite{2002A&A...389..513M} show that only $\sim15\%$ of the gaseous clumps in M16 (dubbed Evaporating Gaseous Globules - EGGs - by \cite{1996AJ....111.2349H}) contain infrared point sources indicative of young stars or protostellar objects. It is evident that the problem of induced star formation is a difficult one requiring much further study.\\

\section{Impact of gravity and cloud structure on the effectiveness of feedback}
The initial conditions of the calculations presented above are clearly very complex. It is obviously of interest to attempt to determine what influence the structure present in our model protoclusters had on their evolution. It was already remarked that the radial density profiles of the models resemble power-laws. Smoothing out over $4\pi$ steradians the mass in spherical shells centred on $r=0$ revealed that all the model clusters could be well represented by density profiles of the form
\begin{eqnarray}
\rho = \left\{\begin{array}{lll}
\rho_{core} & \textrm{for} & r\leq r_{core}\\
\rho_{core}(r_{core}/r)^{-2} & \textrm{for} & r_{cluster}>r >r_{core}
\end{array} \right.
\end{eqnarray} 
where $r_{core}\simeq0.01r_{cluster}$.\\
\indent We used a simple one-dimensional Lagrangian code to investigate the effect of the clouds' structures on their evolution. The code's input physics consisted of fluid dynamics, gravity and photoionisation (note that the self-gravity of the gas was included in the calculation of gravitational forces to give better consistency with our SPH simulations). This code was used to model the evolution of initially-static density profiles with the form of Equation 2 corresponding to the high- and low-density SPH calculations.\\
\indent There are four important lengthscales inherent in this problem. The initial efficacy of photoionisation is determined by the \emph{initial} Str\"omgren radius, $R_{s}$ corresponding to the gas number densities in the cores of the smoothed clouds. The \emph{maximum} efficacy of photoionisation is determined by the equilibrium Str\"omgren radius, $R_{eq}$ at which the HII region would come into pressure equilibrium with the ambient gas if the ambient gas were of density $\rho_{core}$ everywhere. In the absence of gravity, the subsequent evolution of the HII region would be determined purely by the size of $R_{s}$ or $R_{eq}$ relative to the third lengthscale of interest, the cloud core radius, $r_{core}$. The propagation of an ionisation front resulting from the sudden ignition of an ionising source in an initially stationary uniform medium passes through two phases \citep[see e.g.][]{1978ppim.book.....S}. In the very early stages of the development of the HII region, the recombination rate inside the HII region is small compared with the photon output of the source and most of the photons emitted ionise fresh material at the ionisation front. The propagation speed of the front is highly supersonic and the timescale on which the HII region grows is much shorter than any relevant dynamical timescales. The gas remains everywhere approximately motionless in this evolutionary phase, during which the ionisation front is termed an `R-type' front. As the HII region grows, the radiation reaching the ionisation front becomes geometrically diluted and the recombination rate inside the HII region increases, eventually becoming comparable to the photon luminosity of the source. The propagation speed of the ionisation front therefore decreases continually. When the speed of the ionisation front becomes comparable to the sound speed in the HII region, the timescale on which fresh gas is ionised becomes longer than the sound-crossing time of the system and the HII region expands dynamically, driving a strong shock before it. The ionisation front is subsequently known as a `D-type' front and is confined behind the shock front because, in a uniform medium, the shock is always able to sweep up enough material to confine the ionisation front. If the medium in which the ionising source is born is not uniform, but instead possesses a power-law density profile $\rho\sim\rho_{core}^{-w}$, this is not necessarily true. It can be shown that, in cored density profiles in which the density of the envelope falls off more steeply than $w=\frac{3}{2}$ , if the HII region is able to expand beyond the core during either the initial R-type phase or the D-type phase, it will expand to infinite radius \citep{1990ApJ...349..126F}.
\begin{table}
\centering
\begin{tabular}{|l|l|l|l|l|l|}
\hline
Run & $R_{s}$ (cm)& $R_{eq}$ (cm) & $R_{esc}$ (cm) & $r_{core}$ (cm)\\
\hline
UH-1& $2.0\times10^{15}$ & $3.4\times10^{17}$ & $5\times10^{16}$ & $3.8\times 10^{16}$\\
\hline
SL-1& $3.0\times10^{16}$ & $5.1\times10^{18}$ & $ 4\times10^{15}$ & $8.2\times10^{16}$ \\
\hline
\end{tabular}
\caption{Comparison of the important lengthscales in the high- and low-density calculations}

\end{table}

\indent Most of the theoretical work on HII regions completed to date assumes the gas involved to be initially at rest and neglects the effects of the gravitational field of the radiation source. The addition of gravitational forces introduces a fourth lengthscale to the problem, which we term the gravitational escape radius, $R_{esc}$. This is the radius within which the escape velocity due to the gravitational field of the ionising source and of the gas itself exceeds the thermal velocity of the ionised gas. One might expect that an HII region which achieves ionisation equilibrium within this radius will not be able to expand.\\
\indent We list in Table 2 the four lengthscales described above corresponding to the smoothed versions of the high- and low-density SPH calculations. In both cases, the initial Str\"omgren radii of the clouds are inside their core radii, so the HII regions in both calculations are expected to come into ionisation equilibrium and enter their D-type expansion phase before encountering the power-law slope in the density profiles. Also in both cases, the equilibrium Str\"omgren radii lie outside the core radii. In the absence of gravity, one would therefore expect the HII regions in both calculations to expand beyond the cores and to ionise the whole clouds. However, the clouds differ in that, in the high-density case, the initial Str\"omgren radius lies \emph{inside} the gravitational escape radius, whereas in the low-density cloud, the reverse is true. This suggests that the HII region in the high-density cloud may be prevented from expanding, but that in the low-density cloud, the HII region should be able to overflow the core and to then engulf the whole cloud.
\begin{figure}
\centering
\includegraphics[width=0.45\textwidth]{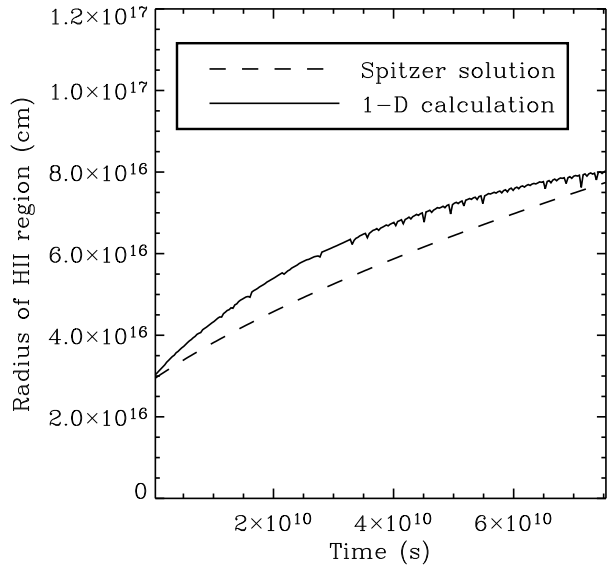}
\caption{Comparison of the evolution of the HII region under the influence of gravity with the Spitzer solution.}
\end{figure}
\indent Both predictions are borne out by the results from our one-dimension calculations. In the high-density case, the HII region had almost no effect on the dynamics of the gas, which rapidly settled into a steady freefall accretion flow. In the low density calculation, the expansion of the ionisation front is similar to the well-known Spitzer solution which would obtain in the absence of gravity in the uniform-density core of the cloud, given by
\begin{eqnarray}
R(t)=R_{s}\left(1+\frac{7}{4}\frac{c_{s}t}{R_{s}}\right)^{\frac{4}{7}},
\end{eqnarray}
where $c_{s}$ is the isothermal sound speed in the ionised gas.\\
\indent The HII region in our calculation expands faster than the Spitzer solution as it accelerates down the density gradient generated by the accretion flow and rapidly reaches the core radius. Once it leaves the core, the HII region will expand rapidly to engulf the whole cloud.\\
\indent We therefore find that gravity prevents the HII region expanding in the case where the Str\"omgren radius is well inside the escape radius, but indirectly accelerates the expansion for $R_{s}>R_{esc}$ by creating a density gradient. \\
\indent The results of our one-dimensional calculations suggest that, if the UH-1 and SH-1 protoclusters investigated in Section 3 had not possessed a high degree of structure, the effect on them of the photoionising radiation from their central massive stars would have been negligible. However, it appears that the structure present in the SL-1 calculation lessened the effectiveness of feedback by preventing the source from ionising all the gas in the system, as the one-dimensional calculation suggests that it should. As one might expect, the `clumpiness' of the gas makes the impact of photoionisation less dependent on the mean gas density.\\
\indent These results have interesting implications for the problem of the energy source required to ionise the Diffuse Interstellar Gas (DIG) in spiral galaxies. The DIG exists as a layer of ionised hydrogen filling large volume fractions of spiral galaxies and accounting for up to $50\%$ of their H$\alpha$ flux (\cite{1996AJ....111.2265F}). The Lyman continuum power required to ionise the DIG is too large to be accounted for by winds or supernovae and the only remaining plausible source is the photoionising radiation from OB stars. In order to produce diffuse ionised gas, this radiation must be able to escape from HII regions. Our results show that this is indeed possible but that the quantity of radiation that can escape an HII region depends strongly on its internal structure. In our high-density simulations, the fractions of flux escaping the clouds towards the ends of the runs was $\sim20\% (\sim1.2\times10^{49}$ s$^{-1})$, whereas the HII region in the corresponding one-dimensional calculation was trapped within the cluster. It was therefore only due to the clumpiness of the gas that any radiation was able to escape the HII region. Conversely, in the low-density SPH simulations, $\sim70\% (\sim1.4\times10^{49}$ s$^{-1})$ of the flux was escaping by the end of the run. In the one-dimensional calculation, the cloud became fully ionised and the fraction of flux escaping would approach unity. The inhomogeneities in the gas therefore limited the escaping flux in the low-density calculations. Although the fluxes should be treated only as estimates, since the boundaries of our clouds were arbitrary, our results suggest that large fluxes of ionising photons can escape from HII regions and are thus a potential means of ionising the DIG. Reynolds in \cite{1997pgh..conf.....L} calculates that $\sim15-20\%$ of the total Lyman continuum from O-stars would be sufficient to ionise the DIG in the Galaxy, which our results demonstrate is feasible.\\
\indent Our calculations also provide an interesting insight into the ultracompact HII region lifetime problem (\cite{1989ApJS...69..831W}). Several authors (e.g. \cite{1995RMxAA..31...39D}, \cite{1996ApJ...469..171G}) have suggested that the expansion of Ultra-Compact HII Regions (UCHIIRs) may be prevented by pressure confinement if the temperatures and number densities of the cores in which the HII regions form are somewhat higher than those typically assumed for molecular clouds or if additional magnetic or turbulent pressures are present. The expansion of the HII regions in our SPH calculations is indeed strongly affected by dynamical pressure, namely the ram-pressure of the accretion flows delivering gas into the vicinity of the radiation source. However, because these ram-pressures are highly anisotropic, they only hinder or prevent expansion of the HII region in some directions, leading to the complex geometry shown in Figure 7. Our one-dimensional calculations also suggest that the gravitational field of the source may be able to prevent HII region expansion.\\
\indent \cite{1996ApJ...456..662K} and \cite{1999ApJ...514..232K} both conducted surveys of UCHIIRs in which they found evidence for extended emission around many sources which are regarded as compact. If it can be shown that such extended emission is directly related to the compact sources, this may alleviate the lifetime problem since it implies that the HII regions \textit{have} expanded as expected. However, the source of the clumps of high-density ionised gas embedded within extended, lower-density HII remains a problem. This topic is discussed by \cite{2000Ap&SS.272..169F}, who suggest that the clumps may be formed by instabilities during the formation or expansion phase of the HII region. We suggest another possibility. The densest ionised gas in our simulations is located at the tips of the filamentary accretion flows and is visible in Figure 7 as the small white region at the centre of the image. Although this very dense material is constantly being accreted, it is also constantly being replenished by the accretion flows, so the small region of very dense ionised gas near the source will persist for as long as the accretion flows.\\
\indent The interaction of HII regions with accretion flows in the presence of gravitational fields is evidently a problem which requires a great deal of further study.\\
\section{Conclusions}
We have undertaken the first ever hydrodynamic simulations of star cluster formation that incorporate feedback from ionising radiation from massive stars. These simulations differ from preceding feedback simulations \citep[e.g.][]{1986MNRAS.221..635T} in that the initial gas density distribution has the highly complex morphology that one would expect to develop in the gravitationally unstable regions of molecular clouds. In particular, the ionising source is located in the cluster core, close to the intersection of a number of dense, nearly radial filaments of inflowing gas. Our chief results are as follows:\\
\indent (1) Radial filaments of gas in the cluster core provide a mechanism for collimating a number of outflows from the vicinity of the massive star. The opening angles, flow velocities and mass-loss rates compare very favourably with those of observed outflows in regions of massive star formation. A more detailed analysis is required in order to assess what fraction of outflows can be collimated by the gaseous environment in this way, compared with the number of sources that require an intrinsic collimation mechanism close to the star.\\
\indent (2) The simulations demonstrate both positive and negative feedback effects. Ionisation reduces the rate of inflow of gas into the cluster core and thus reduces the growth rate of the most massive star compared with control simulations without feedback. On the other hand, feedback also enhances the production of \textit{new} stars due to lateral compression of neutral gas in the walls separating adjacent ionised flows. The latter process is not fully resolved by our calculation, so we cannot at present quantify the relative importance of positive and negative feedback effects.\\
\indent (3) We find that ionisation exacerbates small asymmetries in the initial gas distribution and rapidly generates grossly asymmetric gas structures (see Figures 4, 11 and 16). The impact of ionisation on the cloud (in terms of the volume-filling factor of ionised gas) can be greatly enhanced by the anisotropic gas distribution. The `damage' inflicted by ionisation can readily be assessed from figure 11. By contrast, if the same ionising source is ignited in a cluster whose gas distribution was an azimuthally-averaged rendition of the gas distribution we employ here, we find that the ionising region would be completely trapped within the cluster core (i.e. within $\sim0.02$ pc of the source). However, we also find that the impact of feedback can be decreased by such structure, as shown by comparing our SL-1 run with the corresponding azimuthally-averaged calculation, the latter suggesting that all the gas in the system should be ionised. Instead, the formation of high-density structures protects some of the gas in the system, at least temporarily, from the radiation field, allowing it to remain neutral.\\
\indent (4) We find that although the gas in our simulations absorbs a quantity of thermal and kinetic energy from the ionising source that comfortably exceeds the gravitational binding energy of the cluster, this \textit{is not} a good criterion for assessing whether the cluster will ultimately remain bound or not. The reason for this is simply that most of the energy is acquired by a relatively small fraction of the gas (i.e that involved in the outflows) which is accelerated to a speed that significantly exceeds the cluster escape velocity and carries the energy out of the system.\\
\indent (5) Because of the clumpy nature of the gas in our simulations, we found that significant fractions of our sources' ionising fluxes were able to escape our HII regions entirely, although whether the fraction escaping is enhanced or decreased by the structure of the gas depends on the mean gas density. Our results demonstrate that radiation leaking from HII regions is a plausible means of ionising the Diffuse Ionised Gas in spiral galaxies.\\
\indent The above results suggest that the issue of unbinding clusters by photoionisation, and of the local Star Formation Efficiency in massive star-forming regions needs to be revisited. The fact that massive stars - in simulations and in reality - form in the cluster core where the gas is densest in general inhibits photoionisation feedback, although this effect may be mitigated by stellar wind clearing, which is not included in these simulations. As one would expect, density is the key parameter that controls the efficiency of feedback (contrast Figure 11 with Figure 16). This suggests that there are certain areas of mass-radius parameter space for protoclusters' progenitor clouds which allow the runaway growth of massive stars in the cluster core. Thus - quite apart from the issue of whether massive stars form by coalescence or accretion - the density dependence of feedback provides an additional reason why the most massive stars form in the densest cluster environments.\\

\section*{Acknowledgments}
We gratefully acknowledge the anonymous referee for very helpful comments. JED acknowledges support from a PPARC studentship. CJC gratefully acknowledges support from the Leverhulme Trust in the form of a Phillip Leverhulme prize. The computations reported in Section 3.1 were performed using the UK Astrophysical Fluids Facility (UKAFF).

\bibliography{myrefs}

\label{lastpage}

\end{document}